\documentclass[9pt,twocolumn,twoside]{osajnl}

\journal{ol} 

\setboolean{shortarticle}{true}

\usepackage{subfig}
\usepackage{amsmath}
\usepackage{mathtools}
\usepackage{balance}
\usepackage{ulem}
\usepackage{float}
\usepackage{placeins}
\usepackage{lineno}
\usepackage{caption}
\usepackage{ulem}

\title{Homodyne Time-Domain Acousto-Optic Imaging for Low-Gain Photodetectors}

\author[1,*]{Ahiad Refael Levi}
\author[1,]{Yoav Hazan}
\author[2]{Aner Lev}
\author[2]{Bruno G. Sfez}
\author[1]{Amir Rosenthal}

\affil[1]{Andrew and Erna Viterbi Faculty of Electrical Engineering, Technion – Israel Institute of Technology, Technion City 32000, Haifa, Israel}
\affil[2]{The Israel Center for Advanced Photonics (ICAP), Yavne 81800, Israel}

\affil[*]{sahiadl@campus.technion.ac.il}

\begin{abstract}
Acousto-optics imaging (AOI) is a hybrid imaging modality that  is capable of mapping the light fluence rate in deep tissue by local ultrasound modulation of the diffused photons. 
Since the intensity of the modulated photons is relatively low, AOI systems often rely on high-gain photodetectors, e.g. photomultiplier tubes (PMTs), which limit scalability due to size and cost and may significantly increase the relative shot-noise in the detected signal due to low quantum yields or gain noise. 
In this Letter, we have developed a homodyne AOI scheme in which the modulated photons are amplified by interference with a reference beam, enabling their detection with low-gain photodetectors. 
We experimentally demonstrate our approach with a silicon photodiode, achieving over a 4-fold improvement in SNR in comparison to a PMT-based setup.
\end{abstract}

\setboolean{displaycopyright}{false}

\begin{document}

\maketitle

Deep-tissue optical imaging is generally performed by illuminating the tissue over a large area and using the reemitted diffused light to form an image representative of the optical properties of the tissue \cite{Wang2009}. 
By using tomographic illumination and detection patterns, combined with optimization-based inversion algorithms, depth-resolved imaging may be performed, as been demonstrate in the field of diffuse optical tomography (DOT) \cite{Eggebrecht2014, Durduran2010a}. 
However, purely optical techniques such as DOT are inherently limited in their spatial resolution due light diffusion, mathematically characterized by an ill-conditioned inverse problem.     

Acousto-optics imaging (AOI) \cite{wang2004ultrasound, Doktofsky2020} can improve the resolution of purely optical techniques of deep-tissue imaging by using ultrasound (US). 
In AOI, the tissue is both illuminated with a coherent laser and insonified with an ultrasound transducer, leading to a pressure-induced refractive-index modulation and vibrations of the optical scatterers in the insonified regions \cite{Resink2012}. 
As a results, light that travels through the insonified regions experiences a phase modulation with the same frequency as that of the ultrasound wave.
Because of the high coherence of the laser, the reemitted light exhibits a speckle pattern in which the intensity of each speckle grain is temporally modulated with the ultrasound frequency. 
The modulation depth of the speckle pattern may be measured using various methods \cite{Gunther2017}, enabling a localized detection of light in the tissue. 
AOI may be performed with a focused US beam that scans the imaged region \cite{Leveque-Fort:01} or a set of plane US waves \cite{Laudereau:16}, enabling the visualization of the light fluence rate within the tissue with acoustic resolution.

To optimize the signal-to-noise ratio (SNR) of the measurement, parallel detection of uncorrelated speckle grains is required, which is often performed by digital cameras. 
However, the low time resolution of cameras introduces two difficulties that limit their use \textit{in vivo}. 
First, it complicates the distinction between the effect of US modulation and speckle decorrelation, as both these phenomena are faster than the camera frame-rate \cite{Resink:14}. 
Second, it limits the use of US bursts, for which the acoustic time of flight may be used for depth sectioning. 

In order to overcome speckle decorrelation and enable the use of acoustic bursts, time-domain AOI (TD-AOI) may be used, in which the optical detection is performed with detectors that operate at a higher bandwidth than that of the US transducer \cite{lev2000ultrasound, Granot2001, Lev2002}. 
The immunity of TD-AOI to speckle decorrelation has facilitated \textit{in vivo} demonstrations in small animals \cite{Lev:03-invivo} as well as clinical testing for the diagnosis of osteoporosis  \cite{lev2005}. 
An additional advantage of TD-AOI is its ability to locally quantify blood flow at different depths from the spectral broadening of the AOI signal \cite{Racheli2012, Tsalach:15} --- a capability that has been demonstrated for monitoring cerebral blood flow \cite{balberg2020acousto}.

Because of the low intensity of the US-modulated light, TD-AOI is conventionally performed with photodetectors with a high internal gain, e.g. photomultiplier tubes (PMTs) that use the avalanche effect to multiply the current created by a single photon \cite{Levi20, Lev2002, Tsalach:15}. 
However, despite the high gain, this approach may lead to a lower SNR in the AOI measurement. 
First, the avalanche is an inherently stochastic process in which the gain varies randomly, thus increasing the relative shot noise in the signal. 
Second, in the case of PMT, the relative shot noise is further increased by the low quantum yield of the detector. 
Third, the cost and complexity of high-gain detectors limits their scale-up to multi-element arrays, required for high-SNR operation.

In this work, we have developed a new approach for TD-AOI in which the reemitted light is not detected directly, but is rather interfered with a reference beam in a homodyne configuration. 
The interference leads to an optical amplification of the US-modulated light, enabling its detection with low-gain photodetectors.
While homodyne detection often requires controlling the exact phase difference between the signal and reference arms \cite{Riobo2019}, no such control is required in our scheme due to the stochastic nature of the AOI signal. 
We have experimentally tested our approach with a silicon photodiode (PD), demonstrating over a 4-fold increase in SNR in comparison to a PMT, in agreement with the theoretical prediction. 

Assuming that the detection is performed over time scales faster than the speckle decorrelation time, the field of a single speckle may be represented by 
\begin{equation}
    E_i=E_{bkg,i}e^{j\omega t}+E_{AO,i}e^{j\omega t}e^{j\left(2\pi f_{US}t+\phi_i\right)},
    \label{eq:aoField}
\end{equation}
where $\omega$ is the angular frequency of the light, $f_{US}$ is the US frequency and $\phi_i$ is a random phase distributed: $\phi_i \sim U[0,2\pi]$ that represents the phase difference between the modulated and unmoldulated parts of the field. 
$E_{bkg,i}$ and $E_{AO,i}$ are the amplitudes of the unmodulated and modulated fields, respectively, where it is assumed that $|E_{bkg,i}| \gg |E_{AO,i}|$.

In conventional TD-AOI, the measured optical power consists of spatial integration over $N$ speckles, which are statistically independent and identically distributed, leading to the following expression:
\begin{equation}
\begin{split}
    P^{N} &\propto \sum_{i=1}^N\left|E_{bkg,i}e^{j\omega t} + E_{AO,i}e^{j\omega t}e^{j\left(2\pi f_{US}t + \phi_i\right)}\right|^2 \\
                 &= \sum_{i=1}^N I_{bkg,i} +2\sum_{i=1}^N \sqrt{I_{bkg,i}I_{AO,i}}cos\left(2\pi f_{US}t+\phi_{i}\right).
\end{split}
\label{eq:IN}
\end{equation}

where $I_{bkg,i} = |E_{bkg,i}|^2$ and  $I_{AO,i} = |E_{AO,i}|^2$. 
Since the phase $\phi_i$ is uniformly distributed over $2\pi$, one obtains that $\mathbb{E}[P^{N}]=N\mathbb{E}[I_{bkg,i}]$ and $Var[P^{N}]=N\mathbb{E}[I_{bkg,i}I_{AO,i}]$.
Accordingly, the signal in AOI is often calculated as the standard deviation of $P^{N}$ at the frequency $f_{US}$ and is proportional to $\sqrt{NI_{bkg}I_{AO}}$, where $I_{bkg}=\mathbb{E}[I_{bkg,i}]$ and $I_{AO}=\mathbb{E}[I_{AO,i}]$.

In the proposed homodyne scheme for TD-AOI, the field in Eq. \ref{eq:aoField} is interfered with a reference beam, leading to the following expression for a single speckle grain:
\begin{equation}
    E^{int}_i = {E_{ref}e^{j \omega t} + E_{bkg,i}e^{j\omega t} +  E_{AO,i}e^{j\omega t}e^{j(2\pi f_{US}t + \phi_i)}}.
\end{equation}
Assuming $I_{ref} \gg I_{bkg,i}$, the following expression for the power of N grains is obtained by neglecting $I_{bkg}$:
\begin{equation}
     P^{N}_{hom} \propto N I_{ref} +2I_{ref}\sum_{i=1}^N \sqrt{I_{AO,i}}cos\left(2\pi f_{US}t+\phi_{i}\right).
    \label{eq:INInt}
\end{equation}
Accordingly, the magnitude of the AOI signal in Eq. \ref{eq:INInt}, described by the standard deviation of the expression, is given by $\sqrt{NI_{ref}I_{AO}}$, and the average power is given by $NI_{ref}$.

As can be seen from the above analysis, the use of homodyne detection enables one to amplify the AOI signal since $I_{ref} \gg I_{bkg}$, where the goal is to achieve sufficient amplification such that shot noise becomes the dominant noise factor even when low-gain photodetectors are used. 
In the shot-noise-limited case, the noise is proportional to the square root of the average power, leading to an SNR that is proportional only to $\sqrt{I_{AO}}$ and is independent of all the other parameters in Eq. \ref{eq:IN} and \ref{eq:INInt}. Thus, when examining only inherent noise in the optical signal, homodyne TD-AOI attains the same SNR as conventional TD-AOI. 
Accordingly, the advantage of the homodyne approach is that it enables the use of low-gain photodetectors, with potentially lower cost and noise factors, without the cost of increasing the inherent SNR of the optical signal before detection.     

Assuming shot-noise-limited detection, the SNR at the output of the photodetector is given by \cite{Liu:02}:  
\begin{equation}
    SNR_{det.} = \sqrt{\frac{\eta P}{2\hbar \omega F \Delta f}} ,
    \label{eq:SNRPD}
\end{equation}
where $\eta$ is the quantum efficiency, $P$ is the optical power, $F$ is the noise factor, $\hbar$ is Planck's constant, and $\Delta f$ is the measurement bandwidth. 
The noise factor $F$ is a result of gain fluctuations that occur when the gain is achieved via an avalanche process, and is typically smaller than 1.5 \cite{Liu:02}. 
In the case of a PD, whose gain is provided by a trans-impedance amplifier, the gain may be considered constant, i.e. $F=1$. Accordingly, when shot-noise-limited detection is assumed, the gain in SNR achieved by a PD over a PMT may be expressed by 
\begin{equation}
    G_{SNR}=\sqrt{\frac{\eta_{PD}F_{PMT}}{\eta_{PMT}}}.
\end{equation}     

Using the values $\eta_{PD}=76\%$, $\eta_{PMT}=5\%$, and $F_{PMT}=1.4$, given for the components used, a theoretical SNR gain of $G_{SNR}=4.6$ is obtained.

\begin{figure}[t!]
    \vspace{-0.3cm}
    \centering
    \includegraphics[width=\linewidth]{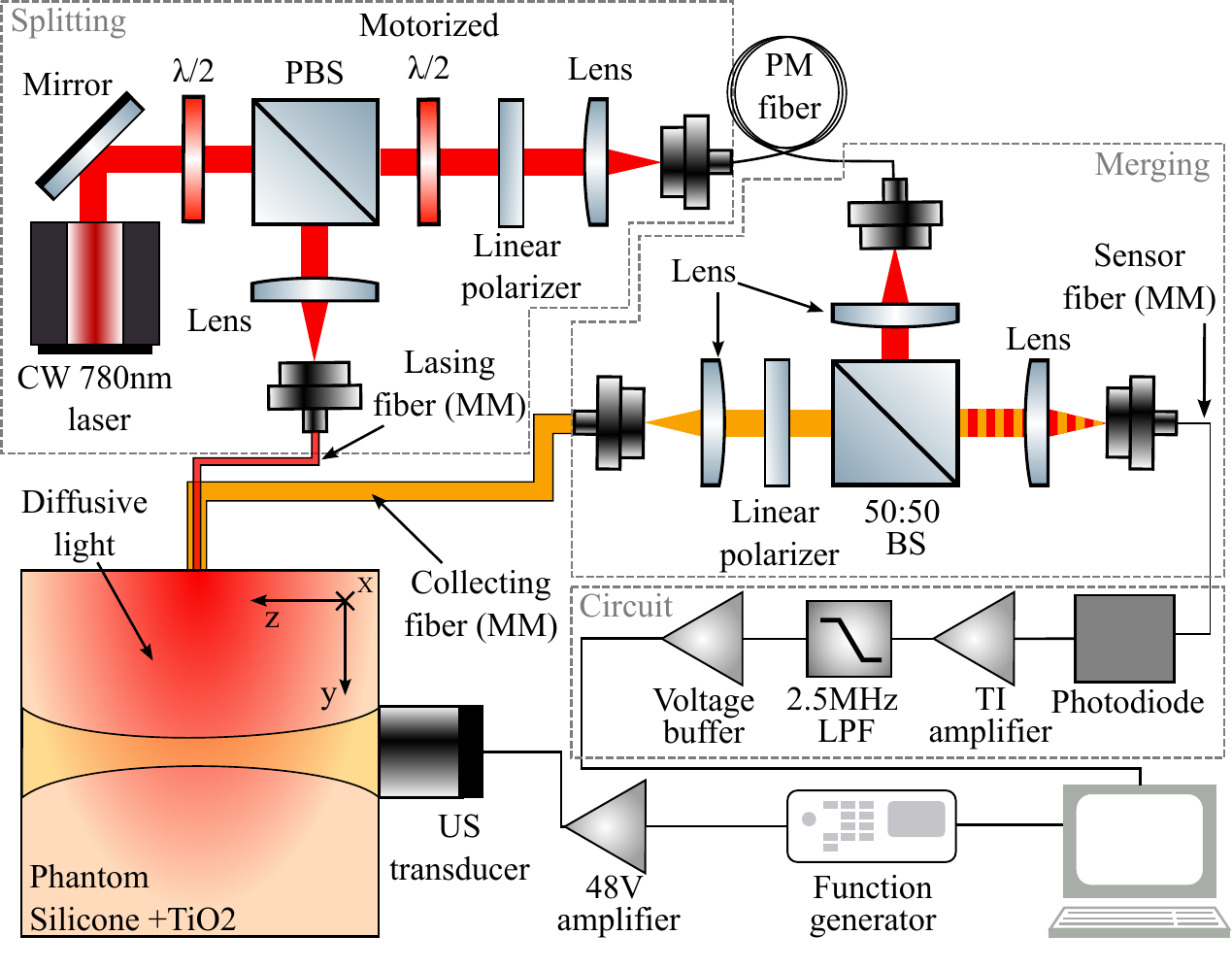}
    \vspace{-0.5cm}
    \caption{AOI system setup. US signal is generated by a function generator, amplified and then fed into a piezoelectric transducer that project it into the phantom. 
    The CW laser beam (780 mn) is split to both illuminating the tissue, and providing the reference beam for later interference.
    The reemitted light collected from the tissue boundary with a MM fiber into the merging optical setup, where it is being polarized and merged with the polarized reference beam using a 50:50 beamsplitter. 
    The interfered light coupled into a MM fiber delivering it onto the PD. 
    The photocurrent from the PD is converted to voltage and filtered before being digitized.}
    \label{fig:systemSetup}
    \vspace{-0.5cm}
\end{figure}

Fig. \ref{fig:systemSetup} shows the experimental setup used in this work to test the performance of homodyne TD-AOI. 
The system's optics is schematically divided into 2 sub-system: (1) splitting module and (2) merging module. 
In the splitting module, a linearly polarized CW laser (DL Pro 780, Toptica) with a linewidth of $50$ KHz and wavelength of $780$ nm is split into 2 branches by two half-wave plates ($ \lambda /2$) and a polarization beam splitter (PBS). 
The first $\lambda/2$ plate angle is set such that most of the optical power is reflected by the PBS into the phantom via a multi-mode fiber with a $62.5 \mu m$ core diameter, 2 m length that delivered $200$ mW to the phantom boundary.
The light transmitted through the PBS is used for the reference beam. 
To optimize the reference beam power, it passes through a second motorized $ \lambda /2$ plate and a polarizer. 
Eventually, the reference beam is coupled to a polarization-maintaining (PM) fiber with a length of 4 m. 

The reemitted light from the phantom, is collected by a second MM fiber with $600 \mu m$ core diameter, 0.39 NA, and 2 m length and delivered to the merging module.
The distance between the two fibers on the phantom boundary was $5$ mm on $X$ axis.
Since the reemitted light is unpolarized, a linear polarizer is used to achieve a polarization that matches that of the reference beam. 
Both beams are merged into a single interfered beam by a 50:50 BS and then coupled into a $200 \mu m$ core diameter fiber. 
The merged light is delivered to a silicon PD (FDS-02, Thorlabs) with a quantum efficiency of $76\%$. 
Both the PD and the PMT were connected to identical custom-made electronics, including a trans-impedance amplifier (Texas Instruments LMH-6626), a $2.5$ MHz passive low pass filter, and a voltage buffer. 
A 14-bit digitizer (ATS-9416, AlazarTech) was used to sample the voltage signal at a frequency of 20 $MHz$.

The technique is tested on a tissue-mimicking phantom made of silicone mixed with $193 nm$ $TiO_2$ particlesforming a reduced scattering coefficient equals to $\mu_s'= 15 cm^{-1}$ and speed of sound of $990$ m/s \cite{Levi20}. 
The ultrasound modulation is generated by a focused piezoelectric transducer (Panametrics, A392S) with diameter of $38.1$ mm, focal length of $9.4 cm$, a Rayleigh length of $3.58$ cm, and a cross-section FWHM of $4.25$ mm. 
The transducer is driven by an arbitrary function generator (Tabor, 8026), amplified to a 48 V peak-to-peak amplitude (.4-1.8-50EU26, SVPA), leading to a peak pressure of approximately $250$ kPa in the acoustic focus. 
The transducer may be used to deliver a single acoustic pulse into the phantom, which modulates different depths at different times, enabling depth-resolved mapping of the AOI signal without scanning by using the time-of-flight principle \cite{Lev:03}. 
In our implementation, we used a coded sequence of pulses, rather than a single pulse, to maximize the acoustic modulation depth \cite{Levi20}. 
Our sequence was based on a code with 251 elements, where each pulse had a single period with a frequency of $f_{US} = 1.25$ MHz, corresponding to an axial resolution of $1.1$ mm, a total imaging depth of $27.6$ cm and sqeuence  repetition rate of $5$ kHz.
In each measurement the signal was measured continuously and averaged over $10^4$ repetitions (a total duration of $2.008$ s). 

To obtain optimal reference power, a preliminary calibration experiment was performed. The transducer was positioned in a fixed coordinates on the $X-Y$ plane and 
$100$ nW of the reemitted light was coupled into the sensor MM fiber (Fig. \ref{fig:systemSetup}). 
In the first part of the experiment, the reference beam was blocked and the reemitted light was directed towards a PMT (R5900U-20-L16, Hamamatsu) with a gain of $2.7 \times 10^5$. 
The PMT signal was used to calculate the baseline SNR of the measurement for conventional TD-AOI.
The measurement was then repeated with the reference beam unblocked, i.e. in a homodyne configuration, for different reference powers. 
For each reference beam intensity, the SNR values were averaged over 30 measurements to minimize the variations between measurements.

\begin{figure}[t!]
    \centering
    \includegraphics[width=0.75\linewidth]{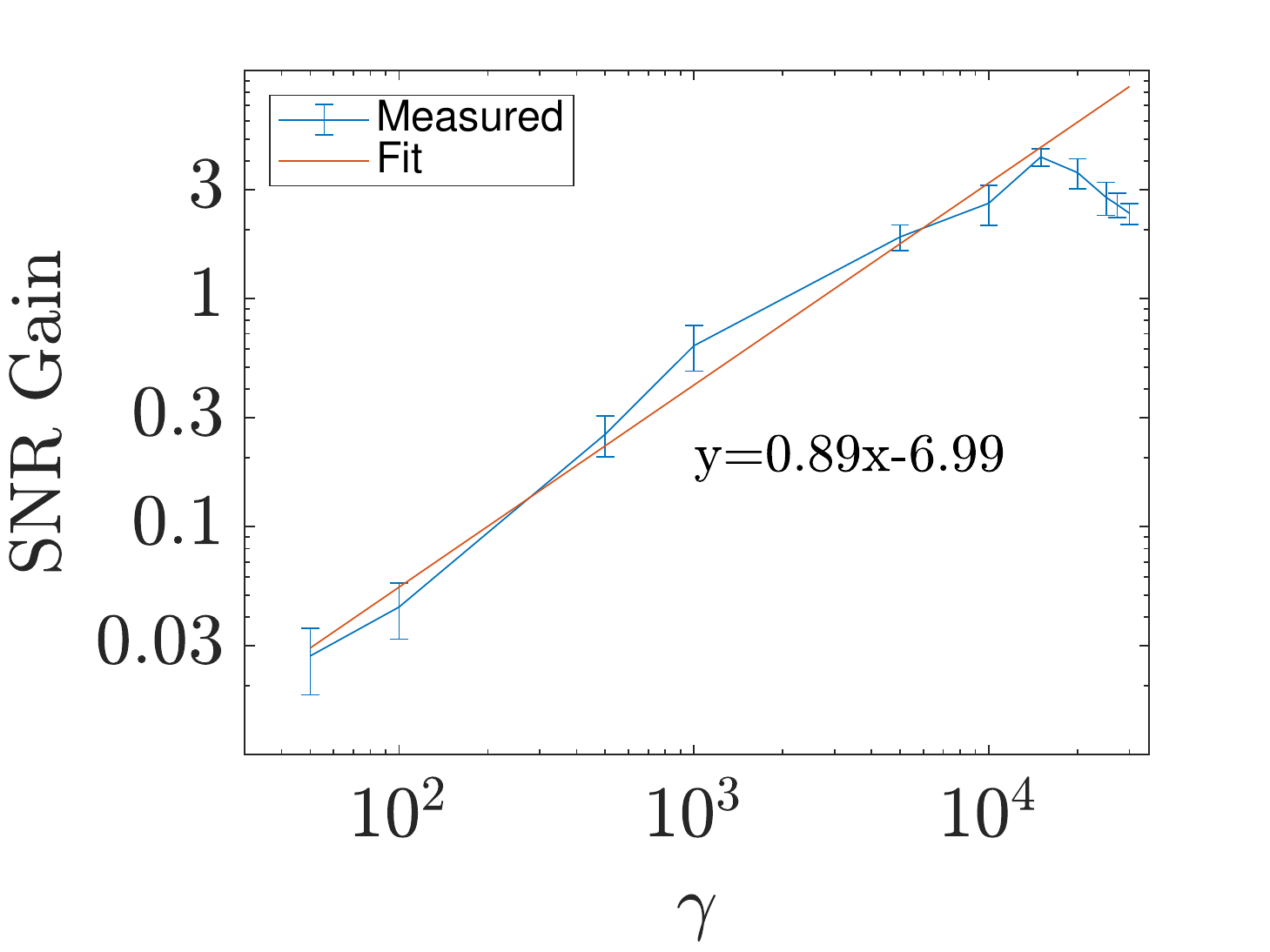}
    \caption{Reference beam intensity calibration. The curve shows the SNR gain obtained with PD vs. the reference beam intensity normalized by the intensity of the reemitted light from the phantom boundary ($\gamma = P_{ref}/P_{re}$). 
    Each point was averaged over 30 repetitive measurements. 
    The curve introduces an exponential rise that ends with a peak SNR gain of 4.16.
    The slow decrease in SNR caused by the saturation of the PD.}
    \vspace{-0.25cm}
    \label{fig:optimizeRef}
\end{figure}

Fig. \ref{fig:optimizeRef} presents the SNR gain of homodyne TD-AOI as a funciton of $\gamma = P_{ref}/P_{re}$, where $P_{ref}$ and $P_{re}$ are the intensities of the reference beam and the reemitted light respectively.
The figure shows that for low values of $\gamma$, the SNR gain is lower than 1, i.e. the homodyne system achieves a lower SNR than conventional TD-AOI. 
The reason for this result is that at low levels of $\gamma$, the detection with PDs is not shot-noise limited, but rather dominated by additive noise from the detector. 
Accordingly, the SNR gain of homodyne-AOI increases approximately linearly with $\gamma$ until a maximum value of 4.17, achieved for $\gamma = 1.5 \times 10^4$, corresponding to reference power of 1.4 mW. 
For higher reference powers, the SNR decreased due to saturation of the PD.  

\begin{figure}[!b]
    \vspace{-0.5cm}
    \centering
    \captionsetup[subfigure]{labelformat=empty}
    \subfloat{
        \includegraphics[height = 3.75cm]{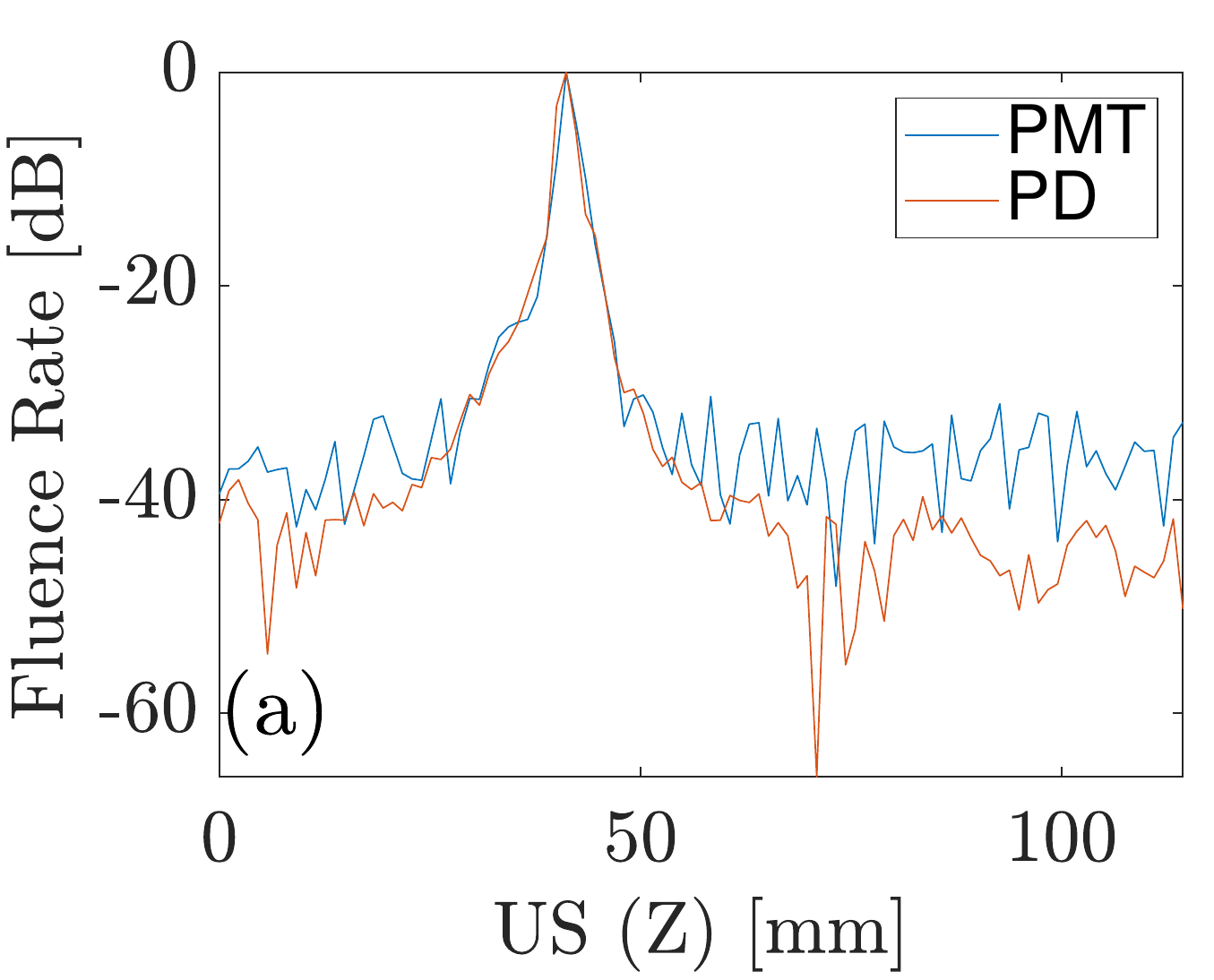}
        \label{fig:usComp}
    } 
    \subfloat{
        \includegraphics[height = 3.75cm]{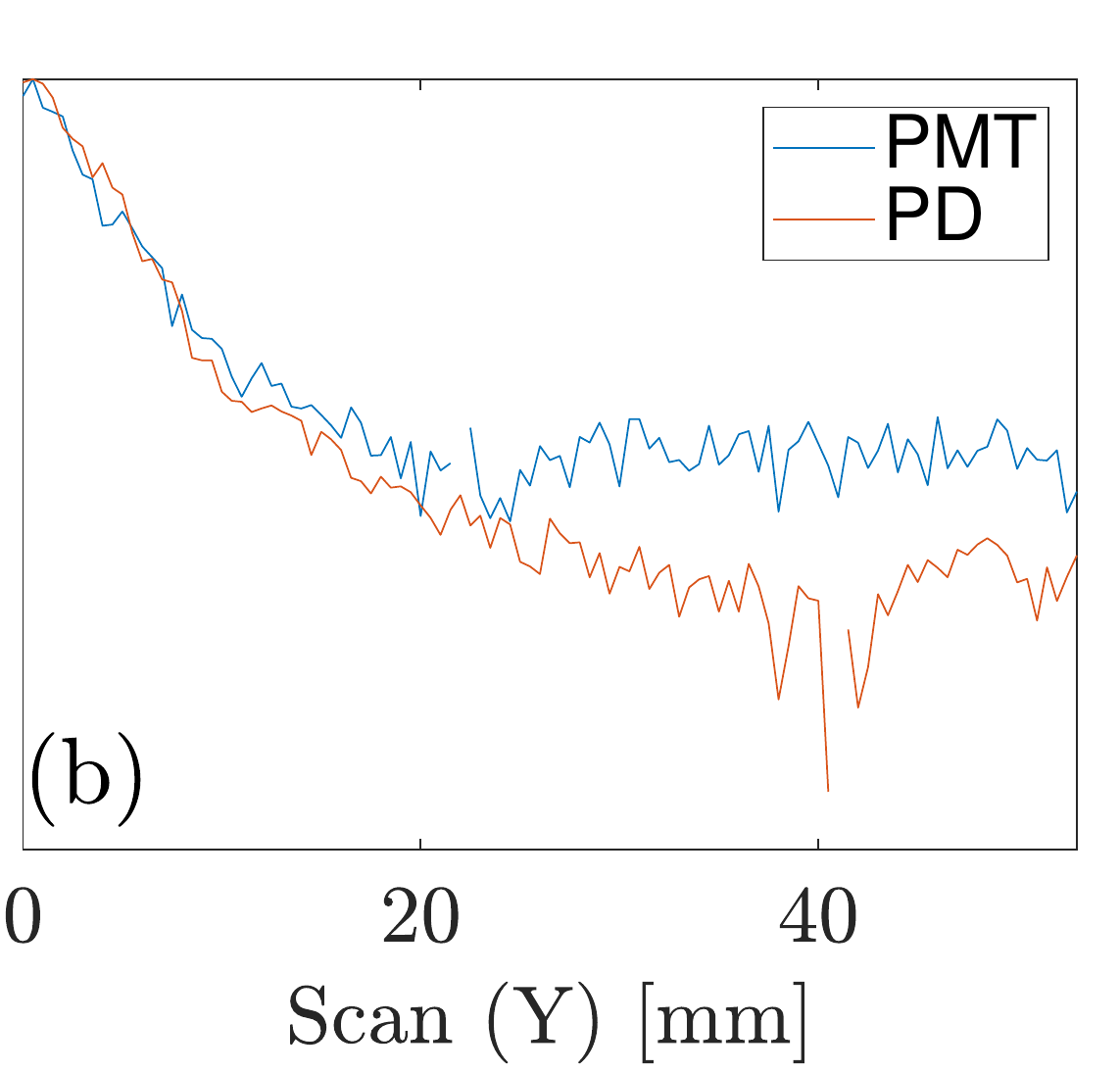}
        \label{fig:scanComp}
    }
    \caption{ Fluence rate profiles measured with conventional TD-AOI using a PMT (blue) and with our homodyne approach using a photodiode (red) along different axes, shown in Fig. \ref{fig:systemSetup}: (a) $z$ axis and (b) $y$ axis. The fluence rate reconstruction from the photodiode-based measurement achieved the same spatial profile for the light fluence rate, but with a noise level that was typically 10 dB lower than that of the PMT-based measurement.}
    \label{fig:profiles}
\end{figure}

Once the reference beam intensity was set to maximal SNR gain, the 1D profile of the fluence rate inside the phantom was mapped in two directions. 
In the $z$ dimension (Fig. \ref{fig:usComp}) the mapping was performed using the time-of-flight principle, whereas in the $y$ dimension (Fig. \ref{fig:scanComp}) the transducer was mechanically scanned with a step size of 0.5 mm. 
The measurement was conducted for both the homodyne and conventional TD-AOI setups, and the results are presented in Fig. \ref{fig:usComp} and \ref{fig:scanComp} for the $z$ and $y$ scans, respectively. 
In both cases, the same spatial profiles were obtained for the AOI signals both the technique.
The asymmetry depicted in the fluence rate shown in Fig \ref{fig:usComp}, which is consistent for both the PMT and the PD measurements, caused by geometric imperfections of tilted fibers with relation to the US axis.
In terms of sensitivity, the noise level in the PMT-based measurement was higher by approximately 10 dB than the one performed with a PD, leading to a difference in penetration depths of over 1 cm in Fig. \ref{fig:scanComp}.

\begin{figure}[t!]
    \centering
    \includegraphics[width=0.75\linewidth]{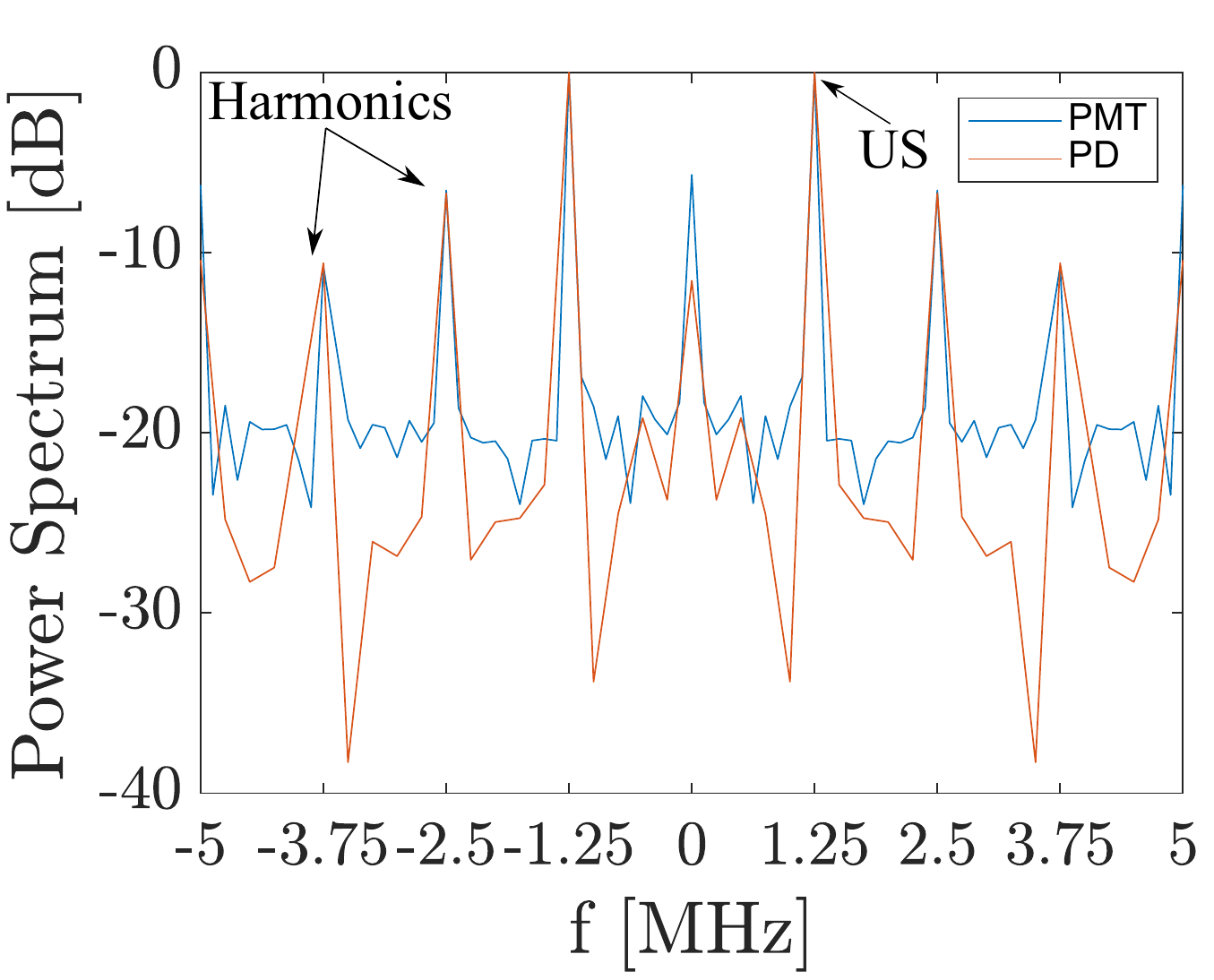}
    \caption{Normalized power spectrum measured in the spatial position in which the AOI was maximal using conventional TD-AOI with a PMT (blue) and our homodyne approach with a photodiode (red). 
    A consistent decrease in noise level measured in both cases in favour of photodiode. 
    The 2nd and 3rd harmonics of the US signal are visible for both techniques.}
    \vspace{-0.25cm}
    \label{fig:fftPeak}
\end{figure}

The results of homodyne and conventional TD-AOI were also compared in the frequency domain. 
Fig. \ref{fig:fftPeak} shows the normalized power spectra of the AOI signals for the spatial positions $y=0$ mm and $z=45$ mm in which the strongest signals were obtained (Fig. \ref{fig:profiles}). 
The figure shows that both techniques led the same spectral behavior, with a strong response at the US frequency $f_{us}=1.25$ MHz, and weaker responses obtained at higher harmonics. 

In conclusion, we have introduced a homodyne detection method for TD-AOI, in which the light reemitted from the tissue is interferred with a reference beam. 
The interference leads to an optical amplification of the AOI signal that enables its detection by low-gain photodetectors. 
In this work, a silicon PD was used in the homodyne scheme, achieving over a 4-fold enhancement in SNR in comparison to conventional TD-AOI performed with a PMT, explained by the higher quantum yield of the PD and its lower noise factor. 
The improved sensitivity of homodyne TD-AOI enabled deeper penetration without distorting the spatial or spectral behavior of the signals. 

In addition to the SNR enhancement demonstrated in this work, our homodyne TD-AOI has the advantage of being compatible with PDs, which are considerably more compact and affordable than PMTs. 
Accordingly, homodyne TD-AOI is more compatible with parallelized detection in which an array of photodetectors simultaneously measure the AOI signal in uncorrelated speckle patterns, facilitating a further improvement in SNR. 

\begin{backmatter}
\bmsection{Funding} This work has received funding from the Ministry of Science, Technology and Space (3-12970) and from the Ollendorff Minerva Center.
\bmsection{Acknowledgments} The authors would like to acknowledge Mr. Moshe Namer for his significant contribution to this work.
\bmsection{Disclosures} The authors declare no conflicts of interest.
\bmsection{Data availability} The data files generated during and/or analyzed during the current study are available from the corresponding author on a reasonable request.
\end{backmatter}

\bibliography{sample}

\begin{thebibliography}{10}
\newcommand{\enquote}[1]{``#1''}

\bibitem{Wang2009}
L.~V. Wang and H.-I. Wu, \emph{{Biomedical Optics}} (John Wiley \& Sons, 2009).

\bibitem{Eggebrecht2014}
A.~T. Eggebrecht, S.~L. Ferradal, A.~Robichaux-Viehoever, M.~S. Hassanpour,
  H.~Dehghani, A.~Z. Snyder, T.~Hershey, and J.~P. Culver, \enquote{{Mapping
  distributed brain function and networks with diffuse optical tomography},}
  {\protect\JournalTitle{Nature Photonics}} \textbf{8}, 448--454 (2014).

\bibitem{Durduran2010a}
T.~Durduran, R.~Choe, W.~B. Baker, and A.~G. Yodh, \enquote{{Diffuse optics for
  tissue monitoring and tomography},} {\protect\JournalTitle{Reports on
  Progress in Physics}} \textbf{73} (2010).

\bibitem{wang2004ultrasound}
L.~V. Wang, \enquote{Ultrasound-mediated biophotonic imaging: a review of
  acousto-optical tomography and photo-acoustic tomography,}
  {\protect\JournalTitle{Disease markers}} \textbf{19}, 123--138 (2004).

\bibitem{Doktofsky2020}
D.~Doktofsky, M.~Rosenfeld, and O.~Katz, \enquote{{Acousto optic imaging beyond
  the acoustic diffraction limit using speckle decorrelation},}
  {\protect\JournalTitle{Communications Physics}} \textbf{3}, 1--8 (2020).

\bibitem{Resink2012}
S.~G. Resink, \enquote{{State-of-the art of acousto-optic sensing and imaging
  of turbid media},} {\protect\JournalTitle{Journal of Biomedical Optics}}
  \textbf{17}, 040901 (2012).

\bibitem{Gunther2017}
J.~Gunther and S.~Andersson-Engels, \enquote{{Review of current methods of
  acousto-optical tomography for biomedical applications},}
  {\protect\JournalTitle{Frontiers of Optoelectronics}} \textbf{10}, 211--238
  (2017).

\bibitem{Leveque-Fort:01}
S.~L\'{e}v\^{e}que-Fort, \enquote{Three-dimensional acousto-optic imaging in
  biological tissues with parallel signal processing,}
  {\protect\JournalTitle{Appl. Opt.}} \textbf{40}, 1029--1036 (2001).

\bibitem{Laudereau:16}
J.-B. Laudereau, A.~A. Grabar, M.~Tanter, J.-L. Gennisson, and F.~Ramaz,
  \enquote{Ultrafast acousto-optic imaging with ultrasonic plane waves,}
  {\protect\JournalTitle{Opt. Express}} \textbf{24}, 3774--3789 (2016).

\bibitem{Resink:14}
S.~Resink, E.~Hondebrink, and W.~Steenbergen, \enquote{Solving the speckle
  decorrelation challenge in acousto-optic sensing using tandem nanosecond
  pulses within the ultrasound period,} {\protect\JournalTitle{Opt. Lett.}}
  \textbf{39}, 6486--6489 (2014).

\bibitem{lev2000ultrasound}
A.~Lev, Z.~Kotler, and B.~G. Sfez, \enquote{{Ultrasound tagged light imaging in
  turbid media in a reflectance geometry},} {\protect\JournalTitle{Optics
  letters}} \textbf{25}, 378--380 (2000).

\bibitem{Granot2001}
E.~Granot, A.~Lev, Z.~Kotler, B.~G. Sfez, and H.~Taitelbaum,
  \enquote{{Detection of inhomogeneities with ultrasound tagging of light},}
  {\protect\JournalTitle{Journal of the Optical Society of America A}}
  \textbf{18}, 1962 (2001).

\bibitem{Lev2002}
A.~Lev and B.~G. Sfez, \enquote{{Direct, noninvasive detection of photon
  density in turbid media},} {\protect\JournalTitle{Optics Letters}}
  \textbf{27}, 473 (2002).

\bibitem{Lev:03-invivo}
A.~Lev and B.~Sfez, \enquote{In vivo demonstration of the ultrasound-modulated
  light technique,} {\protect\JournalTitle{J. Opt. Soc. Am. A}} \textbf{20},
  2347--2354 (2003).

\bibitem{lev2005}
A.~Lev, E.~Rubanov, B.~Sfez, S.~Shany, and J.~Foldes,
  \enquote{Ultrasound-tagged light assessment of osteoporosis,} in
  \emph{Advanced Biomedical and Clinical Diagnostic Systems III,}  vol. 5692
  T.~Vo-Dinh, W.~S.~G. M.D., D.~A.~B. M.D., and G.~E. Cohn, eds., International
  Society for Optics and Photonics (SPIE, 2005), pp. 71 -- 78.

\bibitem{Racheli2012}
N.~Racheli, A.~Ron, Y.~M. M.D., I.~Breskin, G.~Enden, M.~Balberg, and
  R.~Shechter, \enquote{{Non-invasive blood flow measurements using ultrasound
  modulated diffused light},} in \emph{Photons Plus Ultrasound: Imaging and
  Sensing 2012,}  vol. 8223 A.~A. Oraevsky and L.~V. Wang, eds., International
  Society for Optics and Photonics (SPIE, 2012), pp. 438 -- 445.

\bibitem{Tsalach:15}
A.~Tsalach, Z.~Schiffer, E.~Ratner, I.~Breskin, R.~Zeitak, R.~Shechter, and
  M.~Balberg, \enquote{Depth selective acousto-optic flow measurement,}
  {\protect\JournalTitle{Biomed. Opt. Express}} \textbf{6}, 4871--4886 (2015).

\bibitem{balberg2020acousto}
M.~Balberg and R.~Pery-Shechter, \enquote{Acousto-optic cerebral monitoring,}
  in \emph{Handbook of Neurophotonics,}  (CRC Press, 2020), pp. 439--458.

\bibitem{Levi20}
A.~Levi, S.~Monin, E.~Hahamovich, A.~Lev, B.~G. Sfez, and A.~Rosenthal,
  \enquote{Increased snr in acousto-optic imaging via coded ultrasound
  transmission,} {\protect\JournalTitle{Opt. Lett.}} \textbf{45}, 2858--2861
  (2020).

\bibitem{Riobo2019}
L.~Riob{\'{o}}, Y.~Hazan, F.~Veiras, M.~Garea, P.~Sorichetti, and A.~Rosenthal,
  \enquote{{Noise reduction in resonator-based ultrasound sensors by using a CW
  laser and phase detection},} {\protect\JournalTitle{Optics Letters}}
  \textbf{44}, 2677 (2019).

\bibitem{Liu:02}
Z.~Liu and N.~Sugimoto, \enquote{Simulation study for cloud detection with
  space lidars by use of analog detection photomultiplier tubes,}
  {\protect\JournalTitle{Appl. Opt.}} \textbf{41}, 1750--1759 (2002).

\bibitem{Lev:03}
A.~Lev and B.~G. Sfez, \enquote{Pulsed ultrasound-modulated light tomography,}
  {\protect\JournalTitle{Opt. Lett.}} \textbf{28}, 1549--1551 (2003).

\end{thebibliography}

\bibliographyfullrefs{sample}


\ifthenelse{\equal{\journalref}{aop}}{%
\section*{Author Biographies}
\begingroup
\setlength\intextsep{0pt}
\begin{minipage}[t][6.3cm][t]{1.0\textwidth} 
  \begin{wrapfigure}{L}{0.25\textwidth}
    \includegraphics[width=0.25\textwidth]{john_smith.eps}
  \end{wrapfigure}
  \noindent
  {\bfseries John Smith} received his BSc (Mathematics) in 2000 from The University of Maryland. His research interests include lasers and optics.
\end{minipage}
\begin{minipage}{1.0\textwidth}
  \begin{wrapfigure}{L}{0.25\textwidth}
    \includegraphics[width=0.25\textwidth]{alice_smith.eps}
  \end{wrapfigure}
  \noindent
  {\bfseries Alice Smith} also received her BSc (Mathematics) in 2000 from The University of Maryland. Her research interests also include lasers and optics.
\end{minipage}
\endgroup
}{}

\end{document}